\documentclass[12pt]{article}
\pdfoutput=1
\usepackage{jheppub}

\usepackage{amsmath,amssymb,amsfonts}
\usepackage{graphicx}
\usepackage{color}
\usepackage{tikz}
\usepackage{dsfont}

\usepackage{booktabs}
\usepackage{multirow}

\usepackage{subfigure}

\def\cA{{\cal A}}
\def\cB{{\cal B}}
\def\cC{{\cal C}}

\def\cG{{\cal G}}

\def\cM{{\cal M}}

\def\cP{{\cal P}}
\def\cR{{\cal R}}
\def\cN{{\cal N}}

\def\CC{{\mathds{C}}}
\def\RR{{\mathds{R}}}
\def\ZZ{{\mathds{Z}}}

\makeatletter\def\l@subsubsection#1#2{}%
\makeatother

\def\s{\mathfrak{s}}
\def\t{\mathfrak{t}}

\def\Im{\mathop{\rm Im}}
\def\Re{\mathop{\rm Re}}

\def\be{\begin{eqnarray}}
\def\ee{\end{eqnarray}}

\newcommand{\pslash}{\ensuremath\diagup\!\!\!\!\!{+}}

\begin{document}

\title{M-theory curves from warped AdS$_6$ in Type~IIB}
%\title{M-theory perspective on warped AdS$_6$ in Type~IIB}

\author{Justin Kaidi and}
\emailAdd{jkaidi@physics.ucla.edu}
\author{Christoph F.~Uhlemann} 
\emailAdd{uhlemann@physics.ucla.edu}

\affiliation{Mani L.\ Bhaumik Institute for Theoretical Physics\\
Department of Physics and Astronomy\\
University of California, Los Angeles, CA 90095, USA\\[-4mm]}

\abstract{
We establish a close relation between recently constructed AdS$_6$ solutions in Type IIB supergravity, which describe the near-horizon limit of $(p,q)$ 5-brane junctions, and the curves wrapped by M5-branes in the M-theory realization of the 5-brane junctions. This provides a geometric interpretation of various objects appearing in the construction of the Type IIB solutions and a physical interpretation of the regularity conditions. Conversely, the Type IIB solutions provide explicit solutions to the equations defining the M-theory curves associated with $(p,q)$ 5-brane junctions.
}

\maketitle

\section{Introduction and Summary}

One of the remarkable outcomes of string theory is strong evidence for the existence of interacting superconformal field theories (SCFTs) in five and six dimensions. These theories do not admit a conventional Lagrangian description, but they can be realized as low-energy limits of string and M-theory, which allows one to study e.g.\ their moduli spaces and relevant deformations. In many cases, deformations can be found that do admit an effective Lagrangian description, allowing for a match to effective field theory analyses and providing further evidence for the constructions. 

Five-dimensional SCFTs, which are the main concern in this work, can be realized in a variety of ways. First realizations were given in Type IIA on the worldvolume of D4-branes probing a stack of D8-branes and O8$^-$-planes \cite{Seiberg:1996bd,Brandhuber:1999np,Bergman:2012kr}. 
More general classes of theories can be realized in Type IIB on the intersection point of $(p,q)$ five-brane junctions \cite{Aharony:1997ju,Aharony:1997bh,DeWolfe:1999hj}, and in M-theory either on Calabi-Yau threefolds \cite{Morrison:1996xf,Douglas:1996xp,Intriligator:1997pq} or by considering the worldvolume theory of an M5-brane wrapping a holomorphic curve with one compact direction \cite{Kol:1997fv,Brandhuber:1997ua,Aharony:1997bh,Kol:1998cf}.

The AdS/CFT correspondence provides a complementary approach, where the 5d SCFT is identified with a dual string theory on a background with an AdS$_6$ factor. When there exists a brane construction of the 5d SCFT, the dual AdS$_6$ solution is expected to describe the near-horizon geometry of the branes. This is the case for the gravity duals of the 5d SCFTs realized by the D4/D8 system, which have been studied extensively \cite{Ferrara:1998gv,Brandhuber:1999np,Bergman:2012kr,Passias:2018swc,Passias:2012vp,Jafferis:2012iv,Gutperle:2018axv}. More recently, gravity duals have also been constructed for 5d SCFTs realized by $(p,q)$ five-brane junctions in Type IIB \cite{D'Hoker:2016rdq,DHoker:2016ysh,DHoker:2017mds}.\footnote{Previous analyses of the BPS equations can be found in \cite{Apruzzi:2014qva,Kim:2015hya,Kim:2016rhs} and T-duals of the Type IIA solution have been discussed in \cite{Lozano:2012au,Lozano:2013oma}.} Various aspects of the solutions and the dual SCFTs have since been studied holographically \cite{Gutperle:2017tjo,Kaidi:2017bmd,Gutperle:2018wuk}, and comparisons to field theory calculations supporting the proposed dualities have been presented in \cite{Bergman:2018hin,Fluder:2018chf}.
The solutions have also been extended to describe five-brane webs containing mutually local seven-branes \cite{DHoker:2017zwj,Gutperle:2018vdd}, and consistent truncations to 6d $F(4)$ gauged supergravity were constructed in \cite{Hong:2018amk,Malek:2018zcz}.

The geometry of the Type IIB supergravity solutions of  \cite{D'Hoker:2016rdq,DHoker:2016ysh,DHoker:2017mds} is ${\rm AdS}_6 \times {\rm S}^2$ warped over a Riemann surface $\Sigma_{\rm IIB}$, and the solutions are given in terms of a pair of locally holomorphic functions $\cA_\pm$ on $\Sigma_{\rm IIB}$. For the solutions to be physically regular, $\Sigma_{\rm IIB}$ is required to have a boundary and the functions $\cA_\pm$ are required to satisfy certain constraints, to be reviewed below. Along the boundary of $\Sigma_{\rm IIB}$, the differentials $\partial \cA_\pm$ have poles, at which the semi-infinite external five-branes of the associated 5-brane web emerge. The $(p,q)$ charges of the emerging 5-brane are fixed by the residues of $\partial \cA_\pm$. The solutions are completely specified by the choice of Riemann surface $\Sigma_{\rm IIB}$, together with the number of poles and associated residues.

The prominent role of a Riemann surface and holomorphic functions in specifying the Type IIB supergravity solutions may seem reminiscent of the data used by Seiberg and Witten to specify 4d $\cN=2$ theories \cite{Seiberg:1994rs,Witten:1997sc}. Indeed, the same data can be used to specify 5d $\cN=1$ theories engineered by $(p,q)$ 5-brane webs in Type IIB -- that is, such theories may be defined by a holomorphic curve $\Sigma_{\rm M5}$, which contains one compact direction,  together with a holomorphic one-form $\lambda$ on that curve  \cite{Kol:1997fv,Brandhuber:1997ua,Aharony:1997bh,Kol:1998cf}. The physical interpretation is that the 5d $\cN=1$ theory is the worldvolume theory of an M5-brane wrapped on $\Sigma_{\rm M5}$. This suggests that the Riemann surface and holomorphic data characterizing the Type IIB supergravity solutions may be related to the Riemann surface wrapped by the M5-brane in M-theory.

In this paper, we show that this expectation is indeed realized, and explicate the relationship between $\Sigma_{\rm IIB}$ with the locally holomorphic functions $\cA_\pm$ on the one hand, and $\Sigma_{\rm M5}$ with a holomorphic one-form $\lambda$ on the other.
More precisely, we will argue that the locally holomorphic functions $\cA_\pm$ provide an embedding of the doubled Type IIB Riemann surface $\hat\Sigma_{\rm IIB}$ into the flat M-theory geometry, and that this embedded surface {\it is} the surface $\Sigma_{\rm M5}$ wrapped by the M5-brane. The Seiberg-Witten differential $\lambda$ is identified with a locally holomorphic one-form $\cA_+\partial\cA_--\cA_-\partial\cA_+$, which features prominently in the construction of the Type IIB solutions.

This identification between the data defining the Type IIB supergravity solutions and the data used to construct 5d SCFTs in M-theory is useful in a variety of ways. For the Type IIB solutions, it provides a geometric and physical understanding of certain aspects of the construction that are not directly apparent in Type IIB. For example, the physical meaning of the regularity conditions is not immediately apparent in the original formulation. In the M-theory picture, on the other hand, they become the simple condition that the BPS masses associated with the punctures of $\Sigma_{\rm M5}$ vanish - i.e. they enforce conformality of the dual 5d theory. This gives a physical reason for the absence of Type IIB AdS$_6$ solutions with $\Sigma_{\rm IIB}$ being an annulus, or more generally a Riemann surface with multiple boundary components or higher genus. Such solutions would map to M-theory curves describing mass deformations of 5d SCFTs, and are thus not expected to have the full AdS$_6$ isometries. For the solutions with $\Sigma_{\rm IIB}$ being a disc, the identification with the M-theory curve provides independent support for the identification of the solutions with the near-horizon limit of $(p,q)$ 5-brane junctions. 

For the M-theory side, the AdS$_6$ solutions provide explicit solutions to the polynomial equations defining the M-theory curves. We discuss this for a number of explicit classes, where the AdS$_6$ solutions provide simple generating functions for the polynomials defining the curves. This gives a more direct understanding of the pattern of ``binomial edge coefficients,'' discussed in the separate context of brane tilings and their relations to dimer models in \cite{Hanany:2005ve}, and provides a simple way to compute certain multiplicities. We also discuss an interesting relation between the polynomial defining the $T_N$ theory curve and a seemingly unrelated quantity in the field of combinatorics and number theory - namely, the Wendt determinant \cite{Wendt1894,0a45f3e0079b4520b82cdd2698c7a224}. We show that the polynomial defining the M-theory curve for the 5d $T_N$ theories \cite{Benini:2009gi}, evaluated for unit arguments, coincides with the Wendt determinant. 
We leave further exploration to the future, where we certainly expect the connection between Type IIB solutions and M-theory curves to be mutually beneficial.
For example, the M-theory perspective may help identify operators in the SCFTs dual to the Type IIB solutions \cite{Henningson:1997hy,Mikhailov:1997jv}. It may also be useful for generalizing the construction of Type IIB AdS$_6$ solutions with 7-branes \cite{DHoker:2017zwj} to incorporate non-commuting monodromies.

The rest of this paper is organized as follows.  In section \ref{sec:review}, we review the relevant aspects of the Type IIB AdS$_6$ solutions as well as of the M-theory curves. In section \ref{sec:AdS6-M5}, we expand upon the relation between the two pictures and formulate the concrete identification. In section \ref{sec:case-studies}, we verify the proposed identification for five families of supergravity solutions and M-theory curves.

\section{Review: Type IIB \texorpdfstring{AdS$_6$}{AdS6} and M-theory curves} \label{sec:review}
This section contains a review of relevant aspects of the AdS$_6$ solutions in sec.~\ref{sec:ads6}, as well as of the relation between Type IIB 5-brane webs and M5-branes wrapping holomorphic curves in M-theory in sec.~\ref{sec:M5branes}.

\subsection{Warped \texorpdfstring{AdS$_6$}{AdS6} in Type IIB}\label{sec:ads6}
The geometry of the Type IIB AdS$_6$ solutions constructed in \cite{D'Hoker:2016rdq} is a warped product
\begin{align}
 {\rm AdS}_6\times {\rm S}^2\times \Sigma_{\rm IIB}
\end{align}
of AdS$_6$ and S$^2$ over a Riemann surface $\Sigma_{\rm IIB}$. The general solution to the BPS equations is parametrized by two locally holomorphic functions $\cA_\pm$ on $\Sigma_{\rm IIB}$. From these functions a locally holomorphic one-form $d\cB$ on $\Sigma_{\rm IIB}$ is defined,
\begin{align}\label{eq:dB}
 d\cB&=\cA_+d\cA_- - \cA_-d\cA_+~.
\end{align}
The $SL(2,\RR)$ transformations of Type IIB supergravity are induced by a linear action of $SU(1,1)\times\CC$ on the differentials (sec.~5.3 of \cite{D'Hoker:2016rdq}),
\begin{align}\label{eq:SU11-cA}
 \cA_+&\rightarrow u\cA_+ +v \cA_-+c~,&
 \cA_-&\rightarrow \bar v \cA_+ + \bar u\cA_- +\bar c~,
\end{align}
with $|u|^2-|v|^2=1$ and $c\in\CC$. The one-form $d\cB$ is invariant under these transformations.
The shifts parametrized by $c$ leave the supergravity fields invariant, except for a gauge transformation of the two-form field.
The supergravity fields are expressed in terms of $\cA_\pm$, $\cB$, and the composite functions \cite{D'Hoker:2016rdq}
\begin{align}\label{eq:kappa2-G}
 \kappa^2&=-|\partial_w\cA_+|^2+|\partial_w\cA_-|^2~, & 
 \cG&=|\cA_+|^2-|\cA_-|^2+\cB+\bar \cB~,
\end{align}
where $w$ is a local coordinate on $\Sigma$.
Their explicit expressions will not be needed here.

Imposing global regularity conditions constrains the $\cA_\pm$ and requires that $\Sigma_{\rm IIB}$ have non-empty boundary. Physically regular solutions without monodromy were constructed in \cite{DHoker:2016ysh,DHoker:2017mds} for the case in which $\Sigma_{\rm IIB}$ is a disc, or equivalently the upper half-plane. At the boundary of the Riemann surface, $\partial\Sigma_{\rm IIB}$, the spacetime $\rm S^2$ collapses, closing off the ten-dimensional geometry smoothly.
With a complex coordinate $w$ on the upper half-plane, the $\cA_\pm$ are given by
\begin{align}\label{eq:cApm-disc}
 \cA_\pm&=\cA_\pm^0+\sum_{\ell=1}^LZ_\pm^\ell \ln(w-r_\ell)~,
\end{align}
with $\bar Z_\pm^\ell=-Z_\mp^\ell$ and $ \bar\cA_\pm^0=-\cA_\mp^0$. 
The differentials $\partial_w\cA_\pm$ have $L\geq 3$ poles at $w=r_\ell$ on the real line, with residues $Z_\pm^\ell$.
The residues are constructed in terms of a distribution of auxiliary charges and sum to zero by construction.
The locations of the poles are fixed by a set of regularity conditions
\begin{align}
\label{eqn:constr}
 \cA_+^0 Z_-^k - \cA_-^0 Z_+^k 
+ \sum _{\ell \not= k }(Z_+^\ell Z_-^k-Z_+^k Z_-^\ell) \ln |r_\ell - r_k| &=0~,
&k&=1,\cdots,L~.
\end{align}

These physically regular solutions admit a natural identification with $(p,q)$ 5-brane junctions in Type IIB string theory, involving $L$ 5-branes whose charges we denote by $(p_\ell, q_\ell)$ for $\ell=1,..,L$.
At the poles $r_\ell$, the external $(p,q)$ 5-branes of the associated 5-brane junction emerge, with the charges given in terms of the residues by
\begin{align}\label{eq:residue-charge}
 Z_\pm^\ell&=\frac{3}{4}\alpha^\prime (\pm q_\ell+ip_\ell)~,
\end{align}
where a D5-brane corresponds to charge $(\pm 1,0)$ and an NS5-brane to $(0,\pm 1)$ \cite{Bergman:2018hin}.

\subsection{M5-branes on holomorphic curves}\label{sec:M5branes}

Consider a $(p,q)$ 5-brane web in Type IIB in the $(x^5,x^6)$ plane. All 5-branes extend in the field theory directions $x^0,\ldots,x^4$. Compactifying $x^4$ on a circle with radius $R_4$ and T-dualizing leads to Type IIA compactified on the T-dual circle with radius $\tilde R_4=\alpha^\prime/R_4$ and $g_{\rm IIA}=\sqrt{\alpha^\prime}g_{\rm IIB}/R_4$. This is equivalent to M-theory compactified on a torus with coordinates $(x^4,x^{10})$ and $R_{10}=\sqrt{\alpha^\prime}g_{\rm IIA}=g_{\rm IIB} \tilde R_4$. Decompactified Type IIB corresponds to the limit of vanishing volume, $\tilde R_4 R_{10}\rightarrow 0$, with fixed $R_{10}/\tilde R_4$.

In M-theory, the 5-brane web corresponds to a single M5-brane wrapping $x^0,\ldots,x^3$ and a complex curve $\Sigma_{\rm M5}\subset \cM_4$, where $\cM_4 = \RR^2 \times T^2$ is the space spanned by $(x^5,x^6,x^4,x^{10})$.
Using complex coordinates $s$, $t$, defined by
\begin{align}\label{eq:st-def}
 s&=\exp\left(\frac{x^5+i x^4}{\tilde R_4}\right)~, &
 t&=\exp\left(\frac{x^6+i x^{10}}{R_{10}}\right)~,
\end{align}
the curve is an algebraic variety defined by 
\begin{align}\label{eq:Sigma-M5}
 \Sigma_{\rm M5}:\hskip 10mm P(s,t)&=0~.
\end{align}
The polynomial $P(s,t)$ can be constructed in an algorithmic way from the brane web, as will be reviewed shortly, and $\Sigma_{\rm M5}$ is directly related to the Seiberg-Witten curve of the 4d theory obtained by compactifying $x^4$ \cite{Witten:1997sc}.
Supersymmetry requires $\Sigma_{\rm M5}$ to be a calibrated submanifold.
The calibration is given by
\begin{align}
 d\lambda&=d\ln t\wedge d\ln s~,
\end{align}
and the primitive yields the Seiberg-Witten differential, e.g.\ 
\begin{align}\label{eq:lambda}
\lambda&= \frac{dt}{2t}\ln s-\frac{ds}{2s}\ln t~.
\end{align}
The Type IIB $SL(2,\ZZ)$ duality is realized in M-theory as the $SL(2,\ZZ)$ acting on the $(x^4,x^{10})$ torus via
\begin{align}\label{eq:st-SL2Z}
 s&\rightarrow s^a t^b~,&
 t&\rightarrow s^c t^d~,&
 \begin{pmatrix}a & b\\ c& d\end{pmatrix}\in SL(2,\ZZ)~.
\end{align}
The Seiberg-Witten differential in (\ref{eq:lambda}) is invariant under these $SL(2,\ZZ)$ transformations.

\subsubsection{M-theory curves and grid diagrams}
The polynomial $P(s,t)$ defining $\Sigma_{\rm M5}$ is obtained from the grid diagram associated with a given 5-brane web \cite{Aharony:1997bh}. The grid diagram is constructed by placing one vertex in each face of the web and connecting vertices in adjacent faces by a line that crosses the intermediate 5-brane perpendicularly. This gives a convex polygon $\Delta(P)\subset \ZZ^2$.\footnote{The grid diagram is also referred to as the Newton polygon.} One may read off the polynomial $P(s,t)$ from $\Delta(P)$ as follows:
for each point in $\Delta(P)$ with coordinates $(\alpha_i,\beta_i)\in \ZZ^2$, one adds a monomial $s^{\alpha_i}t^{\beta_i}$ with an arbitrary coefficient, resulting in
\begin{align}\label{eq:P-gen}
 P(s,t)&=\sum_{i} c_{i} s^{\alpha_i} t^{\beta_i}~.
\end{align}
Explicit examples will be shown in section \ref{sec:case-studies}.

Now consider one of the asymptotic 5-branes with charges $(p,q)$, in all-ingoing convention.
Supersymmetry requires the slope of this brane in the $(x^5,x^6)$-plane to be
\begin{align}
\frac{\Delta x^6}{\Delta x^5} = \frac{\Im(\tau) q}{p + \Re(\tau) q}~.
\end{align}
This is the condition that there be zero force at the vertices of the web. In M-theory, holomorphicity demands that this constraint be completed to an analogous constraint on $s$ and $t$. The imaginary part of the holomorphic constraint is 
\begin{align}
\frac{\Delta x^{10}}{\Delta x^4} &= \frac{\Im(\tau) q}{p + \Re(\tau) q}~.
\end{align}
Interpreting $\tau$ as the modular parameter of the M-theory torus, this fixes the M5-brane to be oriented along the $(p,q)$ cycle of $T^2$. 

Without loss of generality, we set the asymptotic value of the axio-dilaton scalar to $\tau_{\infty}=i$.\footnote{In M-theory this corresponds to $\tilde R_4=R_{10}$. Expressions for generic values of $\tau_\infty$ are obtained by replacing $x^5\rightarrow \tilde{x}^5  = x^5 - \Re(\tau_\infty)/\Im(\tau_\infty)x^6$, $x^6\rightarrow \tilde{x}^6 = x^6/\Im(\tau_\infty)$ \cite{Aharony:1997bh}.}
The  embedding of the $(p,q)$ 5-brane into the $(x^5,x^6)$-plane is then given by 
\begin{align}\label{eq:pq-5-brane-embedding}
m+(-q x^5+p x^6)T_s&=0~,                                                                                                                                                                                                                                                                                                                             \end{align}
where $m$ corresponds to a mass parameter. 
The projection of the M5-brane curve onto the $(x^5,x^6)$-plane should approach this embedding asymptotically. 
In the $s,t$ coordinates, (\ref{eq:pq-5-brane-embedding}) becomes $\exp(m/ \tilde R_4 T_s)|s|^{-q} |t|^p=1$, while the asymptotic region corresponds to $-p x^5, -qx^6\rightarrow\infty$, or $|s|^{-p},|t|^{-q}\rightarrow \infty$.
In summary, the M-theory curve should behave as
\begin{align}\label{eq:P-bc-1}
 A s^{-q}t^p &\sim 1 ~, & \text{\,\,for\,\,\,\,\, $|s|^{-p},|t|^{-q}\rightarrow \infty$}~,
\end{align}
with $|A|=\exp(m/\tilde R_4 T_s)$.
Requiring that $P(s,t)=0$ exhibits this behavior puts constraints on the coefficients $c_i$.
For a group of $N$ external 5-branes with charges $(p,q)$, the constraint is
\begin{align}
 P(s,t)&\sim \prod_{i=1}^N (A_i t^{p}-s^{q}) & \text{\,\,for\,\,\,\,\, $|s|^{-p},|t|^{-q}\rightarrow \infty$}~.
\end{align}
In the conformal limit, these 5-branes are coincident, and the M5-brane curve is expected to approach this stack of coincident branes. The boundary condition then becomes
\begin{align}
 P(s,t)&\sim (\alpha t^{p}-s^{q})^N~, & \text{\,\,for\,\,\,\,\, $|s|^{-p},|t|^{-q}\rightarrow \infty$}~,
\end{align}
where $\alpha$ is a phase, i.e. $|\alpha|=1$, which encodes the asymptotic behavior of the M-theory curve in the $(x^4,x^{10})$ directions.

\section{M-theory curves from Type IIB \texorpdfstring{AdS$_6$}{AdS6}}\label{sec:AdS6-M5}

In this section we discuss the connection between AdS$_6$ solutions in Type IIB and the holomorphic curves wrapped by M5-branes in M-theory. 
Our main result is a relation between the Riemann surface $\Sigma_{\rm IIB}$ appearing in the supergravity solution and the M-theory curve $\Sigma_{\rm M5}$.
Detailed evidence for the proposed relation will be presented in section \ref{sec:case-studies}.

\subsection{\texorpdfstring{$\cA_\pm$}{A-pm} and algebraic equations}

Before discussing the identification in detail, we rewrite the locally holomorphic functions $\cA_\pm$ in (\ref{eq:cApm-disc}) in a more suggestive way. Using the relation between residues and 5-brane charges (\ref{eq:residue-charge}), as well as the conjugation relations spelled out below (\ref{eq:cApm-disc}), we have
\begin{align}\label{eq:cA-sttilde}
 \cA_\pm&=\frac{3}{4}\alpha^\prime\left(i\ln \s \pm \ln \t\right)~,
\end{align}
where the combinations $\s$ and $\t$ are defined as
\begin{align}\label{eq:st-tilde-def}
 \s&=e^{\Im a}\prod_{\ell=1}^L(w-r_\ell)^{p_\ell}~,
 &
 \t&=e^{\Re a}\prod_{\ell=1}^L(w-r_\ell)^{q_\ell}~,
\end{align}
and we have introduced a constant $a$ defined by $\cA_+^0\equiv\frac{3}{4}\alpha^\prime a$. With these definitions, the locally holomorphic one-form $d\cB$ defined in (\ref{eq:dB}) takes the form
\begin{align}\label{eq:cB-s-t}
 d\cB&=\frac{9}{8}i{\alpha^\prime}^2\left(\frac{d\s}{\s}\ln\t-\frac{d\t}{\t}\ln \s\right)~,
\end{align}
while $\kappa^2$ and $\cG$ of (\ref{eq:kappa2-G}) are given by
\begin{align}\label{eq:kappa-cG-st}
 \kappa^2 dw\wedge d\bar w&=\frac{9}{8i}{\alpha^\prime}^2
 \big(d \ln \s \wedge \overline{ d\ln \t} - d\ln \t \wedge \overline{d\ln \s}\big)~,
 \\
 \cG&=\frac{9}{8i}{\alpha^\prime}^2\left(\overline{\ln\s}\,\ln \t - \ln\s\, \overline{\ln\t}\right)+\cB+\bar\cB~.
\end{align}

The first claim, which we will verify for a number of explicit examples in section \ref{sec:case-studies}, is that the Riemann surface $\Sigma_{\rm IIB}$ with the locally holomorphic functions $\cA_\pm$ provides a solution to equation (\ref{eq:Sigma-M5}) defining the associated M-theory curve, via the identification
\begin{align}\label{eq:st-sttilde}
 s&=\s~, & t&=\t~.
\end{align}
Note that we could in principle allow for arbitrary rescalings of $s$, $t$ in this identification, corresponding to translations of the web -  cf. (\ref{eq:st-def}). As a first consistency check, we note that the $SL(2,\ZZ)$ transformations of $s$, $t$ in (\ref{eq:st-SL2Z}) induce the corresponding transformations of $\cA_\pm$ in (\ref{eq:SU11-cA}) via (\ref{eq:cA-sttilde}) and (\ref{eq:st-sttilde}). Moreover, the constant shifts by $c,\bar c$ in (\ref{eq:SU11-cA}) correspond to translations in $(x^4,x^{10})$ via (\ref{eq:st-def}).

An immediate consequence of this identification is that the holomorphic one-form $d\cB$ in (\ref{eq:cB-s-t})
is directly related to the  Seiberg-Witten differential $\lambda$ in (\ref{eq:lambda}), via
\begin{align}\label{eq:dB-lambda}
 d\cB&=-\frac{9}{4}i{\alpha^\prime}^2 \lambda~.
\end{align}

\subsection{Global structure}

We have claimed that the functions $\cA_\pm$ on the Riemann surface $\Sigma_{\rm IIB}$ provide a solution to the equation defining the M-theory curve, $\Sigma_{\rm M5}$. We now address this identification at the global level.
The relation (\ref{eq:cA-sttilde}) with (\ref{eq:st-sttilde}) and (\ref{eq:st-def}) in fact suggests a more direct identification of $\cA_\pm$ with the coordinates in M-theory as follows,
\begin{align}\label{eq:cApm-M4}
 \frac{x^{5}+i x^4}{\tilde R_4}&=-\frac{2i}{3\alpha^\prime}\left(\cA_++\cA_-\right)~,
 &
 \frac{x^6+i x^{10}}{R_{10}}&=\frac{2}{3\alpha^\prime}\left(\cA_+-\cA_-\right)~.
\end{align}
That is, the functions $\cA_\pm$ provide an embedding of $\Sigma_{\rm IIB}$ into the four-dimensional space $\cM_4=\RR^2\times T^2$ spanned by the M-theory coordinates $(x^5,x^6,x^4,x^{10})$.
An apparent challenge to a direct identification of $\Sigma_{\rm IIB}$ and $\Sigma_{\rm M5}$ is the fact that, being a disc or the upper half-plane, $\Sigma_{\rm IIB}$ has a boundary, while $\Sigma_{\rm M5}$ does not.
We note that 
\begin{align}
 \bar\cA_\pm -\cA_\mp&=2\pi i \sum_{k=1}^L\Theta(r_k-w)  Z_\mp^k ~, & w\in\partial\Sigma_{\rm IIB}~,
\end{align}
with $\Theta$ the Heaviside function.
Consequently, for integer charges $p_k$, $q_k$,
\begin{align}\label{eq:dSigma-image}
 \frac{x^{10}}{R_{10}}=\frac{x^4}{\tilde R_4}=0 \mod \pi \qquad \forall w\in\partial\Sigma_{\rm IIB}~.
\end{align}
Thus, the segments of the boundary of $\Sigma_{\rm IIB}$ in between poles are mapped to curves in planes of constant $x^4$ and $x^{10}$. 
The embedding of $\Sigma_{\rm IIB}$ into $\cM_4$ is illustrated in Figures~\ref{fig:T1-curve} and \ref{fig:Plus-1-curve} for the $T_1$ and $+_{1,1}$ solutions, respectively.

A natural interpretation for the boundary in $\Sigma_{\rm IIB}$ can be obtained as follows. We recall that the regularity conditions in Type IIB supergravity have two branches of solutions (sec.~5.4 of \cite{D'Hoker:2016rdq}),
\begin{align}
 \cR_+&: \quad \lbrace \kappa^2>0 ~, \quad \cG>0\rbrace~,
 &
 \cR_-&: \quad \lbrace \kappa^2<0 ~, \quad \cG<0\rbrace~.
\end{align}
These two branches are mapped into one another by complex conjugation. 
The regular solutions discussed above with $\Sigma_{\rm IIB}$ being the upper half-plane realize the branch $\cR_+$.
For any such regular solution in the upper half-plane, the extension of the $\cA_\pm$ into the lower half-plane provides an equivalent regular solution, realizing the second branch of regularity conditions $\cR_-$.
The two solutions are separated at the boundary of $\Sigma_{\rm IIB}$, where $\kappa^2=\cG=0$.

Since the 10d spacetime in Type IIB is closed off smoothly at $\partial\Sigma_{\rm IIB}$ by the collapsing S$^2$, the solutions in the upper and lower half-planes are two realizations of equivalent Type IIB solutions. But for the identification of $\Sigma_{\rm IIB}$ with the M-theory curve, it is natural to consider the full, doubled, Riemann surface $\hat\Sigma_{\rm IIB}$.\footnote{ In fact, the construction of regular solutions in \cite{DHoker:2017mds} employed an auxiliary electrostatics potential, in which the doubled Riemann surface $\hat\Sigma_{\rm IIB}$ already played a crucial role.} The precise relation we propose is then
\begin{align}
 \Sigma_{\rm M5}: \qquad \hat\Sigma_{\rm IIB} \  \xrightarrow{\  \cA_\pm \ } \ \cM_4=\RR^2\times T^2~.
\end{align}
That is, the embedding of the doubled Type IIB Riemann surface $\hat\Sigma_{\rm IIB}$ into the four-dimensional part of the M-theory geometry, with the embedding functions given by $\cA_\pm$ via (\ref{eq:cApm-M4}), {\it is} the M-theory curve $\Sigma_{\rm M5}$.

The doubled Type IIB Riemann surface $\hat\Sigma_{\rm IIB}$ is a closed surface with punctures at the poles $r_\ell$.
Suppose we encircle one of the poles $r_\ell$. Then $\ln(w-r_\ell)\rightarrow \ln(w-r_\ell)+2\pi i$, and consequently
\begin{align}
 \cA_+ \pm \cA_- &\rightarrow \cA_+ \pm \cA_- + 2\pi i\left(Z_+^\ell\pm Z_-^\ell\right)
 ~.
\end{align}
With the identifications (\ref{eq:residue-charge}) and (\ref{eq:cApm-M4}), this means that
\begin{align}
 x^4&\rightarrow x^4 + 2\pi  \tilde R_4 q_\ell~, & x^{10}&\rightarrow x^{10}+2\pi  R_{10} p_\ell~.
\end{align}
This is indeed the desired behavior: the $(p,q)$ 5-brane charges become the winding numbers of the M5-brane, with the winding on the M-theory circle $x^{10}$ encoding the D5 charge and the winding on $x^4$ encoding the NS5 charge. 
This furthermore implies that the curve defined by the embedding (\ref{eq:cApm-M4}) is smooth across the boundary of $\Sigma_{\rm IIB}$, despite the fact that the $\cA_\pm$ are not single-valued in the doubled Riemann surface $\hat\Sigma_{\rm IIB}$ (noting that the differentials $\partial_w\cA_\pm$ are single-valued on $\hat\Sigma_{\rm IIB}$).
That is, since
\begin{align}
 \overline{\cA_\pm(\bar w)}&=-\cA_\mp(w) + \frac{3}{2}\alpha^\prime i\pi  k~, & k\in\ZZ~,
\end{align}
mapping from the upper half-plane $\Sigma_{\rm IIB}$ into the lower half-plane of $\hat\Sigma_{\rm IIB}$ induces the following map on the M-theory curve,
\begin{align}
 w&\mapsto \bar w\,: & x^4&\mapsto -x^4 \ \mod 2\pi \tilde R_4~,
\nonumber \\
 & & x^{10}&\mapsto -x^{10}\mod 2\pi R_{10}~.\label{eq:SigmaIIB-to-cc}
\end{align}
Then due to (\ref{eq:dSigma-image}), the boundary of $\Sigma_{\rm IIB}$ is mapped to fixed points of this action on the torus.

\subsection{Type IIB regularity conditions}\label{sec:IIB-reg}

The asymptotic behavior of the M5-brane curve is constrained by the conditions (\ref{eq:P-bc-1}).
We will now discuss how this behavior is realized by the identification (\ref{eq:st-sttilde}), and obtain a geometric perspective on the Type IIB regularity conditions (\ref{eqn:constr}). Consider the limit in which 
\begin{align}
 w&\rightarrow r_k~.
\end{align}
With the explicit expressions in (\ref{eq:st-tilde-def}), we find that in this limit
\begin{align}
 |\s|^{-p_k}, |\t|^{-q_k} &\rightarrow \infty~,
\end{align}
corresponding to the asymptotic region where 5-branes with charges $(p_k,q_k)$ are, as expected.
Furthermore, in this limit the explicit expressions in (\ref{eq:st-tilde-def}) give
\begin{align}
 \s^{-q_k}\t^{p_k}&=e^{p_k\Re(a)-q_k\Im(a)}\prod_{\ell\neq k}(r_k-r_\ell)^{q_\ell p_k-p_\ell q_k}~,
\end{align}
which is finite, as required by (\ref{eq:P-bc-1}). As seen from (\ref{eq:P-bc-1}), the mass parameter associated with the external 5-branes is given by
\begin{align}
 -m_k^2&=\ln\left|\s^{-q_k}\t^{p_k}\right|^2
 \nonumber\\
 &=
 2p_k\Re(a)-2q_k\Im(a)+\sum_{\ell\neq k}\left(q_\ell p_k-p_\ell q_k\right)\ln|r_k-r_\ell|^2~.
\end{align}
Using the identification of the residues with the 5-brane charges (\ref{eq:residue-charge}), as well as the definition of the constant $a$ below (\ref{eq:st-tilde-def}), the Type IIB regularity conditions in (\ref{eqn:constr}) are precisely the statement that $m_k^2=0$ for all $k$.
The Type IIB regularity conditions are therefore interpreted from the M-theory perspective as the requirement that the 5-branes within each group of like-charged external 5-branes are coincident, with the associated mass parameter vanishing.

The identification of $d\cB$ with the Seiberg-Witten differential allows for an additional physical interpretation of the regularity conditions (\ref{eqn:constr}) from the 4d perspective.
Of the $L$ conditions in (\ref{eqn:constr}) only $L-1$ are independent, due to the fact that the $Z_\pm^\ell$ sum to zero by construction, implementing charge conservation at the 5-brane junction.
These conditions may be formulated more concisely in the upper half-plane as
\begin{align}\label{eq:reg-dB}
 \int_{C_k} d\cB +{\rm c.c.}&=0~, &
 k&=1,\ldots,L~,
\end{align}
where $C_k$ denotes a curve connecting two points on the boundary $\partial\Sigma_{\rm IIB}$  to either side of the pole $r_k$. In this formulation, charge conservation amounts to the fact that the sum of the cycles $C_k$ is contractible. 

In the doubled surface $\hat \Sigma_{\rm IIB}$, the addition of the complex conjugate on the left hand side in (\ref{eq:reg-dB}) can be implemented by closing the contour $C_k$ in the lower half-plane, such that the pole is encircled completely. Denoting by $\hat C_k$ a closed contour around the pole $p_k$ in $\hat \Sigma_{\rm IIB}$, the regularity conditions become
\begin{align}\label{eq:reg-dB-2}
 \int_{\hat C_k} d\cB&=0~, &  k&=1,\ldots,L~.
\end{align}
With the identification of $\hat \Sigma_{\rm IIB}$ as the Seiberg-Witten curve of the 5d theory compactified on $x^4$,  and of $d\cB$ as the Seiberg-Witten differential via (\ref{eq:dB-lambda}), the regularity conditions (\ref{eq:reg-dB-2}) again become the statement that the BPS masses associated with the punctures vanish.

\subsection{\texorpdfstring{$\Sigma_{\rm IIB}$}{Sigma-IIB} of general topology}\label{sec:multiple-boundaries}

The identification of $\hat \Sigma_{\rm IIB}$ with the M-theory curve $\Sigma_{\rm M5}$ gives an interesting perspective on potential AdS$_6$ solutions in Type IIB where $\Sigma_{\rm IIB}$ is a Riemann surface with multiple boundary components or higher genus. From the Type IIB perspective, it is not \textit{a priori} clear whether such solutions should exist. The construction used in \cite{DHoker:2017mds} of imposing the global regularity conditions on the general local solution to the BPS equations and reducing them to a finite number of constraints in principle works for Riemann surfaces of arbitrary topology. This was spelled out explicitly in sec.~6 of \cite{DHoker:2017mds}. But solutions to these constraints were only found for the upper half-plane. For the annulus, an explicit search was conducted, but no solutions were found.

From the perspective of the associated M-theory curve, assuming that the identification of $\hat \Sigma_{\rm IIB}$ with $\Sigma_{\rm M5}$ extends to $\Sigma_{\rm IIB}$ of more general topology, $\Sigma_{\rm IIB}$ with multiple boundaries or higher genus would correspond to M-theory curves $\Sigma_{\rm M5}$ of higher genus. Such curves are associated to 5-brane webs with open faces, i.e.\ mass deformations. These webs describe renormalization group flows, as opposed to renormalization group fixed points, and are therefore not expected to have an AdS$_6$ dual. This gives a physical interpretation for the absence of annulus solutions in Type IIB, and suggests more generally the absence of AdS$_6$ solutions for Riemann surfaces with multiple boundary components or higher genus.

\section{Case studies}\label{sec:case-studies}
In this section, we verify the relation between the Type IIB AdS$_6$ solutions and M-theory curves discussed in sec.~\ref{sec:AdS6-M5} for a number of explicit examples.

\subsection{\texorpdfstring{$T_N$}{T-N} solutions}

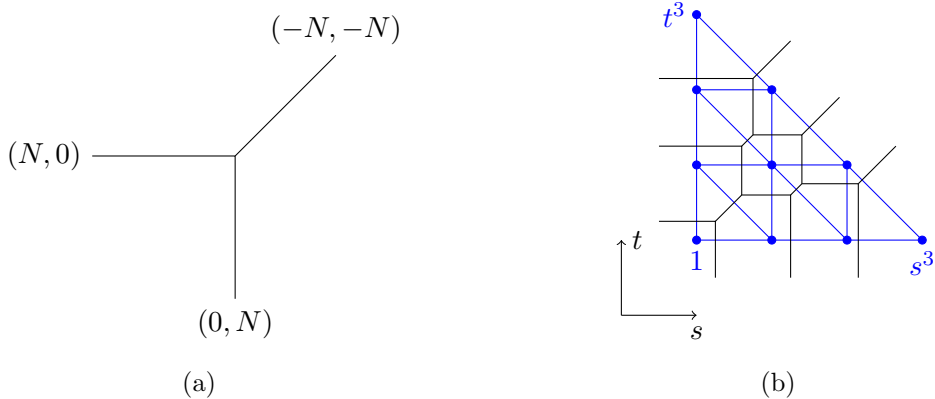
\begin{figure}
\centering
\subfigure[][]{\label{fig:TN-junction}
 \begin{tikzpicture}[scale=0.95]
   \draw (-1,0) node [anchor=east] {\small $(N,0)$} -- (1,0) -- (1,-2) node [anchor=north] {\small $(0,N)$};
   \draw (1,0) -- + (1.41,1.41) node [anchor=south] {\small $(-N,-N)$};
 \end{tikzpicture}
}\hskip 25mm
\subfigure[][]{\label{fig:T3-grid}
 \begin{tikzpicture}
   \foreach \i in {0,...,3}{
   \foreach \j in {0,...,\i}{
     \draw[fill,blue] (3-\i,\j) circle (1.5pt);
   }
   }
   
    \draw[blue] (0,0) -- (3,0) node [anchor=north] {\small $s^3$}-- (0,3)node [anchor=east] {\small $t^3$} -- (0,0) node [anchor=north] {\small $1$};
    \draw[blue] (0,2) -- (2,0);
    \draw[blue] (0,1) -- (1,0);
    \draw[blue] (0,2) -- (1,2);
    \draw[blue] (0,1) -- (2,1);
    \draw[blue] (1,0) -- (1,2);
    \draw[blue] (2,0) -- (2,1);

   \draw (-0.5,0.25) -- (0.25,0.25) -- (0.25,-0.5);
   \draw (0.25,0.25) -- (0.6,0.6) -- (1.25,0.6) -- (1.25,-0.5);
   \draw (0.6,0.6) -- (0.6,1.25) -- (-0.5,1.25);
   
   \draw (0.6,1.25) -- +(0.15,0.15) -- (0.75,2.15) -- (-0.5,2.15);
   \draw (0.75,2.15) -- +(0.5,0.5);
   
   \draw (1.25,0.6) -- +(0.15,0.15) -- (2.15,0.75) -- (2.15,-0.5);
   \draw (2.15,0.75) -- +(0.5,0.5);
   
   \draw (0.75,1.4) -- (1.4,1.4) -- (1.4,0.75);
   \draw (1.4,1.4) -- +(0.5,0.5);
   
    \draw[->] (-1,-1) -- (-1,0) node [anchor=west] {\small $t$};
    \draw[->] (-1,-1) -- (0,-1) node [anchor=north] {\small $s$};
 \end{tikzpicture}
}

\caption{
Left: the 5-brane junction describing the $T_N$ SCFTs with charge assignments in ingoing convention.
Right: brane web and grid diagram for a mass deformation of the $T_3$ theory. Some examples of the monomials associated to the grid points are shown. }
\end{figure}

As a first example we discuss the 5d $T_N$ theories \cite{Benini:2009gi}. These are realized by triple junctions of $N$ D5, $N$ NS5, and $N$ $(1,1)$ 5-branes (fig.~\ref{fig:TN-junction}). 
The polynomial $P(s,t)$, obtained from the grid diagram (fig.~\ref{fig:T3-grid}), is given by
\begin{align}\label{eq:Pst-TN}
 P(s,t)&=\sum_{i=0}^N\sum_{j=0}^{N-i}c_{i,j} s^i t^j~.
\end{align}
The boundary conditions, in the conformal limit, are 
\begin{align}
 \text{$s,t\rightarrow \infty$}:& & P(s,t)&\sim \sum_{k=0}^Nc_{k,N-k} s^k t^{N-k} &\stackrel{!}{\sim}   & \ \ (s-\alpha_1 t)^N~,
 \nonumber\\
 \text{$s$ finite, $t\rightarrow 0$}:& & P(s,t)&\sim\sum_{k=0}^N c_{k,0} s^k &\stackrel{!}{\sim} & \ \ (s-\alpha_2)^N~, 
 \nonumber\\
 \text{$t$ finite, $s\rightarrow 0$}:& & P(s,t)&\sim\sum_{k=0}^N c_{0,k} t^k &\stackrel{!}{\sim} &\ \ (1-\alpha_3 t)^N~,
\label{eq:T-N-bc}
\end{align}
with $|\alpha_i|=1$.
This fixes the coefficients $c_{k,N-k}$, $c_{k,0}$ and $c_{0,k}$ for $k=0,\ldots,N$ to be binomial. The remaining coefficients encode Coulomb branch parameters. 
Without loss of generality, we fix $c_{0,0}=1$. Then for $N=1$, one finds
\begin{align}\label{eq:P-TN-final}
 P_{T_1}(s,t)&=1-\alpha_2^{-1}s-\alpha_3t~.
\end{align}
Consistency of the boundary conditions requires $\alpha_1=\alpha_2\alpha_3$. The remaining freedom in $\alpha_2$, $\alpha_3$ corresponds to translations in the compact directions.

\begin{figure}
\centering
 \begin{tikzpicture}
  \node at (0,0) {\includegraphics[width=0.5\linewidth]{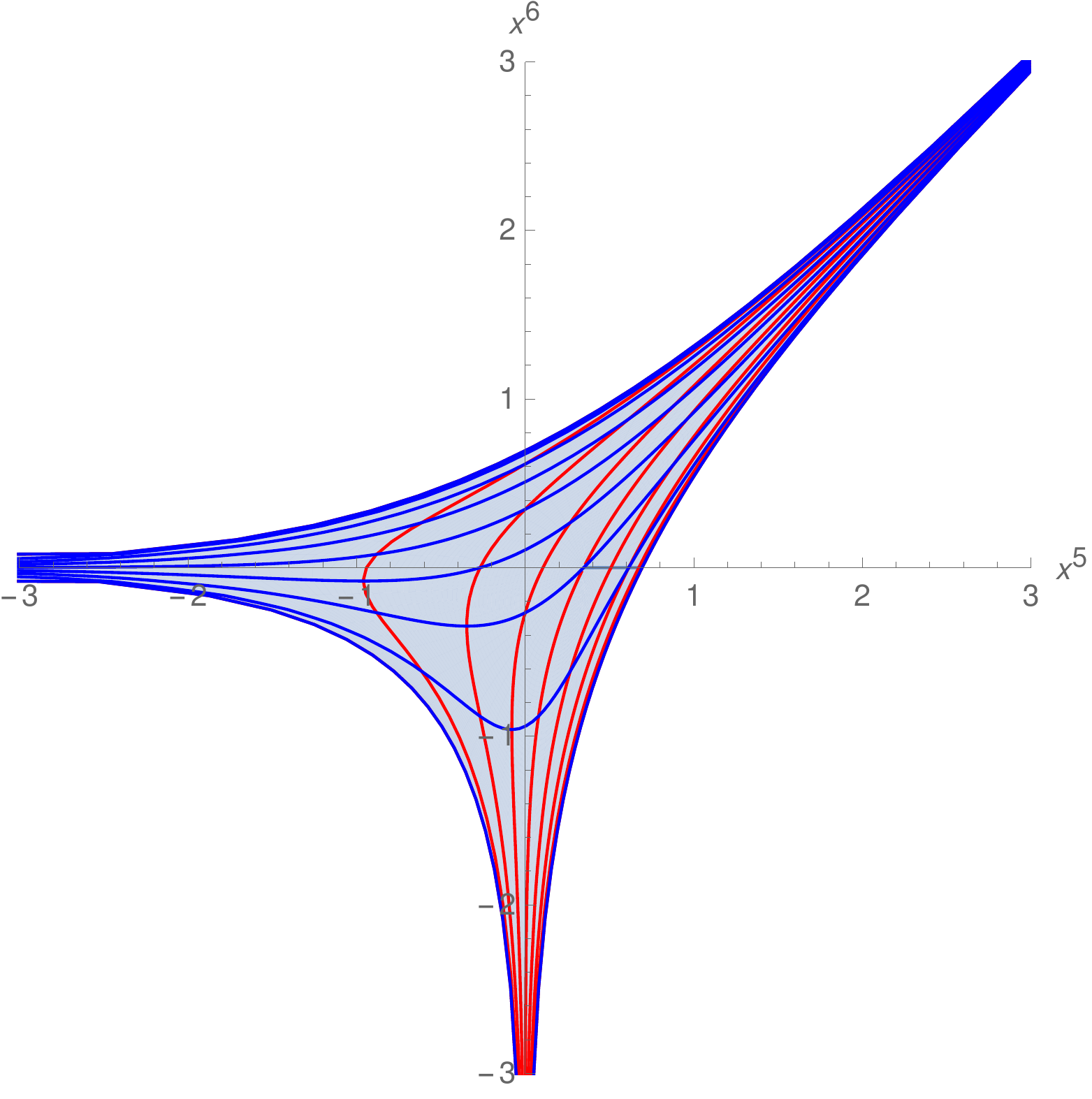}};
  \node at (0.9,-3.5) {$r_1=1$};
  \node at (-3.5,-0.6) {$r_2=0$};
  \node at (4,3) {$r_3=-1$};
  
  \node[rotate=60] at (1.4,-0.0) {\small $(x^4,x^{10})=(0,0)$};
  \node[rotate=25] at (-0.6,1.2) {\small $(x^4,x^{10})=(\pi,\pi)$};
  \node[rotate=-55] at (-1.5,-1.7) {\small $(x^4,x^{10})=(0,\pi)$};
 \end{tikzpicture}
 \caption{$T_1$ curve with $\tilde R_4=R_{10}=1$ obtained by embedding $\Sigma_{\rm IIB}$ into $\cM_4$ via (\ref{eq:cApm-M4}). The poles $r_\ell$ on $\Sigma_{\rm IIB}$ correspond to the external 5-branes in the asymptotic regions as indicated.  The segments of the boundary $\partial\Sigma_{\rm IIB}$ in between poles are mapped to the outer curves connecting the asymptotic regions, with values of $x^4$, $x^{10}$ as indicated. The blue curves correspond to constant $x^{4}$, the red curves to constant $x^{10}$. Both are positive for $w\in\Sigma_{\rm IIB}$. The embedding of the second half of $\hat\Sigma_{\rm IIB}$, with $w$ in the lower half-plane, is obtained by reversing the signs of $x^4$ and $x^{10}$ (\ref{eq:SigmaIIB-to-cc}).\label{fig:T1-curve}}
\end{figure}
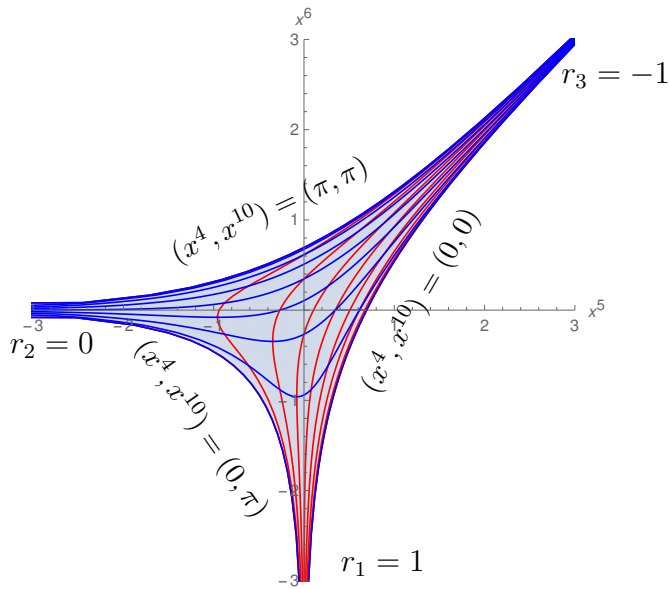

The Type IIB supergravity solutions corresponding to triple junctions of D5, NS5, and $(1,1)$ 5-branes were discussed in detail in \cite{Bergman:2018hin,Fluder:2018chf}, including comparisons of holographic results to field theory computations. The functions $\cA_\pm$ are given by (sec.~4.3 of \cite{Bergman:2018hin})
\begin{align}
 \cA_\pm &=\frac{3}{4}\alpha^\prime N \left[\pm\ln(w-1)+i\ln (2w) - (i\pm 1)\ln(w+1)\right]~.
\end{align}
This realizes the $T_N$ charges in all-ingoing convention.
Via (\ref{eq:cA-sttilde}) this yields
\begin{align}\label{eq:st-tt-TN}
 \s&=\left(\frac{2w}{1+w}\right)^N~, & \t&=\left(\frac{w-1}{w+1}\right)^N~.
\end{align}
For $N=1$, these solve (\ref{eq:P-TN-final}) with $\alpha_2=-\alpha_3=1$ via $s=\s$ and $t=\t$ . 
More generally, $\s$ and $\t$ satisfy
\begin{align}\label{eq:cP-TN}
 0&=\cP_{T_N}(\s,\t)~, & \cP_{T_N}(\s,\t)&\equiv1-\s^{1/N}+\t^{1/N}~.
\end{align}
Solving this equation for either $\s$ in terms of $\t$ or $\t$ in terms of $\s$ yields $N$ branches of solutions. These are realized in (\ref{eq:st-tt-TN}) by the fact that solving for $w$ in terms of $\s$ or  $\t$ yields $N$ branches of solutions. Evaluating the expression for the remaining one of $\s$ or $\t$ for these $w$ gives $N$ branches for $\s$ in terms of $\t$ and $\t$ in terms of $\s$.

Eq.~(\ref{eq:cP-TN}) can be converted to a polynomial equation $\tilde P_{T_N}(\s,\t)=0$ with the same roots.
The result is
\begin{align}\label{eq:Ptilde-TN}
 0&=\tilde P_{T_N}(\s,\t)~,
 &
 \tilde P_{T_N}(\s,\t)&\equiv\prod_{n=0}^{N-1}\prod_{m=0}^{N-1} \cP_{T_1}\left(e^{\frac{2 \pi i n}{N}} \s^{\frac{1}{N}}, e^{\frac{2 \pi i m}{N}} \t^{\frac{1}{N}} \right)~.
\end{align}
This is indeed a polynomial in $\s$ and $\t$ for each $N$, where each term has combined degree at most $N$, as in (\ref{eq:Pst-TN}); all fractional powers of $\s$ and $\t$ drop out.
This shows that the subspace in $\cM_4$ defined by (\ref{eq:cApm-M4}) is indeed an algebraic variety.
That the polynomial satisfies the boundary conditions spelled out in (\ref{eq:T-N-bc}) for general $N$ can be verified directly by inspecting $\cP_{T_N}$ in (\ref{eq:cP-TN}). It also follows from the general discussion in sec.~\ref{sec:IIB-reg}, which showed that $\s$ and $\t$ extracted from regular supergravity solutions automatically realize the appropriate asymptotic behavior. Some explicit forms of the coefficients $\tilde c_{ij}$ of $\tilde P_{T_N}(\s,\t)=\sum_{ij} \tilde c_{ij}\s^i\t^j$ for small $N$ are
\begin{align}\label{eq:TN-smallN}
\tilde c_{ij}^{T_2}&=\begin{pmatrix}
 1 & -2 & 1 \\
 -2 & -2  \\
 1  \\
\end{pmatrix}
&
\tilde c_{ij}^{T_3}&=\begin{pmatrix}
 1 & 3 & 3 & 1 \\
 -3 & 21 & -3 \\
 3 & 3  \\
 -1  \\
\end{pmatrix}~.
\end{align}
The coefficients which are not fixed by the boundary conditions (\ref{eq:T-N-bc}) are tuned to specific values, corresponding to the origin of the Coulomb branch. This is the expected result for the curve extracted from a Type IIB supergravity solution with an AdS$_6$ factor, describing the conformally invariant vacuum state.

We now  discuss the mapping of the Type IIB Riemann surface $\Sigma_{\rm IIB}$ to the M-theory curve. With the identification of $\s$, $\t$ given in (\ref{eq:st-tt-TN}) with $s$, $t$ and their relation (\ref{eq:st-def}) to the M-theory coordinates $(x^5,x^6,x^4,x^{10})$ on $\cM_4=\RR^2\times T^2$, we obtain the embedding of $\hat\Sigma_{\rm IIB}$ into $\cM_4$ as
\begin{align}
 x^5+ix^{4}&=\tilde R_4 N\ln\left(\frac{2w}{1+w}\right)~,
 &
 x^6+ix^{10}&=R_{10} N \ln\left(\frac{w-1}{w+1}\right)~.
\end{align}
The poles at $r_1$, $r_2$, $r_3$ correspond to the NS5, D5, and $(1,1)$ 5-branes, respectively. 
The geometry of the curve for $N=1$ is illustrated in fig.~\ref{fig:T1-curve}.
The curve for generic $N$ is obtained by a simple rescaling.

We note that eq.~(\ref{eq:Ptilde-TN}) is precisely the formula quoted in (3.13) of \cite{Hanany:2005ve}, which made use of earlier results in \cite{Kenyon:2003uj}. The context of that result was a proposed correspondence between brane tilings and dimer models. Though we have not been considering brane tilings in the current work, the curves wrapped by the NS5-branes in the brane tiling construction are of the same form as the curves being wrapped by the M5-brane here. In the current context, the formula of \cite{Hanany:2005ve} appears more naturally in the form (\ref{eq:Pst-TN}), coming directly from the warped AdS$_6$ solutions. The pattern of binomial coefficients on the edges (cf.~(\ref{eq:TN-smallN})), which was traced back in \cite{Hanany:2005ve} to the expression (\ref{eq:Ptilde-TN}), implements the boundary conditions on the curve as discussed in sec.~\ref{sec:IIB-reg}. 

We also note an interesting relation between the polynomial defining the $T_N$ theory curve and a seemingly unrelated quantity in the field of combinatorics and number theory. Namely, this is the Wendt determinant \cite{Wendt1894,0a45f3e0079b4520b82cdd2698c7a224}, given by
\begin{align}
 W_n&=\prod_{j=0}^{m-1}\left(\left(1+\zeta^j_m\right)^m-1\right)~,
\end{align}
where $\zeta_m$ is a primitive $m$-th root of unity. To make the relation to the polynomial $\tilde P_{T_N}(\s,\t)$ transparent, we note the alternative expression
\begin{align}
\tilde P_{T_N}(\s,\t)&=\prod_{n=0}^{N-1}\left(\left(1+e^{\frac{2\pi i n}{N}}\t^{1/N}\right)^N-\s\right)~. 
\end{align}
This expression shows that the Wendt determinant $W_n$ is obtained by evaluating the polynomial for $\s=\t=1$,
\begin{align}
 W_n&=\tilde P_{T_n}(1,1)~.
\end{align}
The first terms in the sequence are given by
\begin{align}
 W_1&=1~, & W_2&=-3~, & W_3&=28~, & W_4&=-375~, & W_5&=3751~, & W_6&=0~.
\end{align}
The relation of the Wendt determinant to circulant matrices with all binomial coefficients may provide an interesting perspective on the conformal invariance of the curve. We leave further investigation of this relation to the future.

For each theory obtained by wrapping an M5-brane on a holomorphic curve, there is an alternative interpretation as M-theory on a (singular) Calabi-Yau threefold. In the particular case of rank 1 SCFTs with toric realizations (i.e. theories with grid diagrams with a single internal dot), this threefold is a complex cone over $\mathds{F}^0$ or a del Pezzo surface dP$_n$, $n \leq 3$ \cite{Morrison:1996xf,Douglas:1996xp,Intriligator:1997pq}. This may be seen by interpreting the brane web as the toric skeleton defining the geometry \cite{Leung:1997tw}. In the case of the $T_1$ theory, the corresponding Calabi-Yau threefold is simply $\CC^3$. The higher rank $T_N$ theories correspond to orbifolds of $\CC^3$, i.e. $\CC^3 / (\ZZ_N \times \ZZ_N)$ with the orbifold action given by  \cite{Hanany:2005ve}
\begin{align}
\label{orbifoldaction}
(z_1,z_2,z_3) &\mapsto (\lambda z_1, z_2, \lambda^{-1} z_3)~,& \lambda^N &=1~,
\nonumber\\
(z_1,z_2,z_3) &\mapsto ( z_1, \nu z_2, \nu^{-1} z_3)~,& \nu^N &=1~.
\end{align}

\subsection{\texorpdfstring{$Y_N$}{Y-N} solutions}

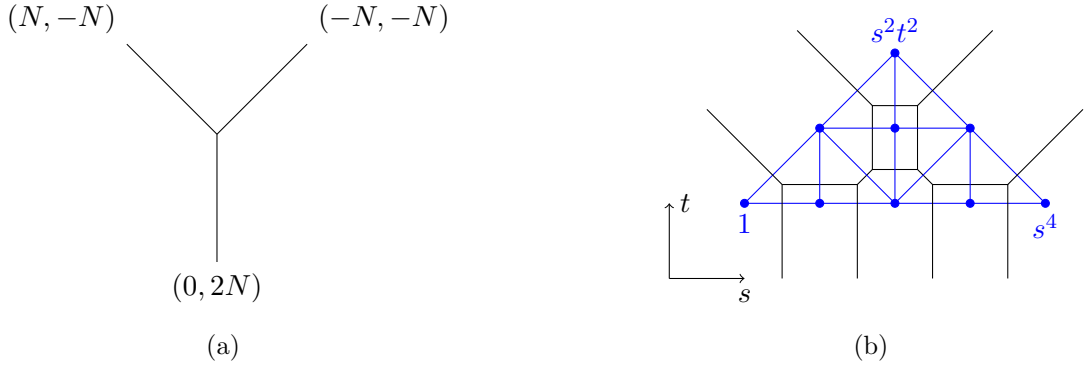
\begin{figure}
\centering
\subfigure[][]{\label{fig:YN-junction}
 \begin{tikzpicture}[scale=0.85]
   \draw (-0.41,1.41) node [anchor=south east] {\small $(N,-N)$} -- (1,0) -- (1,-2) node [anchor=north] {\small $(0,2N)$};
   \draw (1,0) -- + (1.41,1.41) node [anchor=south west] {\small $(-N,-N)$};
 \end{tikzpicture}
}\hskip 25mm
\subfigure[][]{\label{fig:Y2-grid}
 \begin{tikzpicture}
   \foreach \i in {-1,...,1}{  
    \draw[fill,blue] (\i,0) circle (1.5pt);
    \draw[fill,blue] (\i,1) circle (1.5pt);    
   }
   \draw[fill,blue] (-2,0) circle (1.5pt);
   \draw[fill,blue] (2,0) circle (1.5pt);
   
   \draw[fill,blue] (0,2) circle (1.5pt);
   \draw[blue] (-2,0)node [anchor=north] {\small $1$} -- (2,0) node [anchor=north] {\small $s^4$}-- (0,2) node [anchor=south] {\small $s^2 t^2$} -- (-2,0);

   \draw (-1.5,-1) -- (-1.5,0.25) -- (-2.5,1.25);
   \draw (-1.5,0.25) -- (-0.5,0.25) -- (-0.5,-1);
   \draw (-0.5,0.25) -- (-0.3,0.45) -- (0.3,0.45) -- (0.5,0.25) -- (0.5,-1);
   \draw (-0.3,0.45) -- (-0.3,1.3) -- (0.3,1.3) -- (0.3,0.45);
   \draw (-0.3,1.3) -- (-1.3,2.3);
   \draw (0.3,1.3) -- (1.3,2.3);
   \draw (0.5,0.25) -- (1.5,0.25) -- (1.5,-1);
   \draw (1.5,0.25) -- (2.5,1.25);
   
   \draw[blue] (0,2) -- (0,0) -- (1,1) -- (-1,1) -- (0,0);
   \draw[blue] (1,1) -- (1,0);
   \draw[blue] (-1,1) -- (-1,0);
   
   \draw[->] (-3,-1) -- (-3,0) node [anchor=west] {\small $t$};
   \draw[->] (-3,-1) -- (-2,-1) node [anchor=north] {\small $s$};
 \end{tikzpicture}
}

\caption{
Left: the 5-brane junction describing the $Y_N$ SCFTs.
Right: brane web and grid diagram for a mass deformation of the $Y_2$ theory.
}
\end{figure}

As a next example we discuss the closely related $Y_N$ junctions, which are triple junctions of $N$ $(1,1)$ 5-branes, $N$ $(-1,1)$ 5-branes, and $2N$ D5-branes (fig.~\ref{fig:YN-junction}). Although generally different from the $T_N$ junctions, at the level of supergravity the solutions corresponding to the $Y_N$ theories are related to the $T_N$ solutions by an $SL(2,\RR)$ transformation combined with a rescaling of the charges (sec.~4.3 of \cite{Bergman:2018hin}). This leads to simple relations between the large-$N$ limits of the two theories. The curves are likewise closely related, as we will discuss now.

We start with the supergravity picture in this case, and compare to the construction of the curve via the grid diagram associated with the brane web at the end. The functions $\cA_\pm$ are given by
\begin{align}
 \cA_\pm&=\frac{3}{4}\alpha^\prime N\left[(i\mp 1)\ln (w+1)\pm 2\ln (4w)-(i\pm 1)\ln(w-1)\right]~,
\end{align}
from which we extract, via (\ref{eq:cA-sttilde}),
\begin{align}
 \s&=\left(\frac{w+1}{w-1}\right)^N~,
 &
 \t&=\left(\frac{4w^2}{w^2-1}\right)^N~.
\end{align}
They satisfy
\begin{align}\label{eq:cP-YN}
 0&=\cP_{Y_N}(\s,\t)~, & \cP_{Y_N}(\s,\t)&=1+\s^{1/N}-(\s \t)^{1/(2N)}~.
\end{align}

This can be understood from the result for the $T_N$ solution as follows. We first note that $\s$, $\t$ for the $Y_N$ solution are related to $\s$, $\t$ for the $T_N$ solution by
\begin{align}\label{eq:TN-to-YN}
 \s_{Y_N}&=\t_{T_N}^{-1}~, & \t_{Y_N}&=\s_{T_N}^2\t_{T_N}^{-1}~.
\end{align}
This may be interpreted as the $Y_N$ solution being obtained from the $T_N$ solution by an $SL(2,\RR)$ transformation with $a=0$, $c=-1/b=-1/d=\sqrt{2}$, acting as in (\ref{eq:st-SL2Z}), combined with a charge rescaling $N\rightarrow\sqrt{2}N$.
As a consequence of (\ref{eq:TN-to-YN}), we have
\begin{align}
 \cP_{Y_N}(\s,\t)&=\s^{1/N}\cP_{T_N}\left(\sqrt{\frac{\t}{\s}},\frac{1}{\s}\right)~.
\end{align}

We now compare to the polynomial equation obtained from the grid diagram of the $Y_N$ junctions. A sample grid diagram is shown in fig.~\ref{fig:Y2-grid}, and the resulting polynomial takes the form
\begin{align}\label{eq:P-YN}
 P(s,t)&=\sum_{i=0}^{2N}\sum_{j=0}^{N-|N-i|}c_{i,j}s^i t^j~.
\end{align}
The boundary conditions in the conformal limit demand that the coefficients on the edges be binomial. More precisely, the requirements are 
\begin{align}
 P(s,t)\big\vert_{s,t\rightarrow\infty}&\stackrel{!}{\sim}  s^N(s-\alpha_1 t)^N~,
 &
 P(s,t)\big\vert_{s\rightarrow 0, t\rightarrow\infty}&\stackrel{!}{\sim}  s^N\left(t-\frac{\alpha_2}{s}\right)^N~,
 \nonumber\\
 P(s,t)\big\vert_{\text{$s$ finite}, t\rightarrow 0}&\stackrel{!}{\sim}  (s-\alpha_3)^{2N}~.
 \label{eq:Y-N-bc}
\end{align}
Consistency of the boundary conditions requires $\alpha_1 \alpha_2 = \alpha_3^2$.

Eq.~(\ref{eq:cP-YN}), which is satisfied by $\s$ and $\t$ obtained from the supergravity solution, may again be converted to a polynomial equation, $0=\tilde P_{Y_N}(\s,\t)$, as follows. Eq.~(\ref{eq:cP-YN}) for $N=1$ is equivalent to
\begin{align}
 0&=\tilde P_{Y_1}(\s,\t) & \tilde P_{Y_1}(\s,\t)&=(\s+1)^2-\s \t~.
\end{align}
For higher $N\geq 2$, 
\begin{align}
\label{eq:Y-polynomial}
  \tilde P_{Y_N}(\s,\t)&\equiv\prod_{n=0}^{N-1}\prod_{m=0}^{N-1} \tilde P_{Y_1}\left(e^{\frac{2 \pi i n}{N}} \s^{\frac{1}{N}}, e^{\frac{2 \pi i m}{N}} \t^{\frac{1}{N}} \right)~.
\end{align}
This is again a polynomial in $\s$ and $\t$, and takes precisely the form in (\ref{eq:P-YN}). Moreover, the edge coefficients are binomial, reflecting the fact that the curve obtained from the supergravity solution automatically satisfies the correct boundary conditions. Some explicit forms for small $N$ are
\begin{align}
 \tilde c_{ij}^{Y_2}&=\begin{pmatrix}
 1  \\
 -4 & -2  \\
 6 & -12 & 1 \\
 -4 & -2  \\
 1 \\
\end{pmatrix}~,
&
\tilde c_{ij}^{Y_3}&=
\begin{pmatrix}
 1  \\
 6 & -3\\
 15 & 150 & 3  \\
 20 & -423 & 60 & -1 \\
 15 & 150 & 3  \\
 6 & -3  \\
 1  \\
\end{pmatrix}~.
\end{align}
As before, the coefficients corresponding to Coulomb branch deformations are tuned to particular values for the conformally invariant vacuum state. The supergravity solution again provides an explicit solution to the equation defining the M-theory curve, with $\cA_\pm$ providing the embedding as discussed in sec.~\ref{sec:AdS6-M5}.

The $Y_1$ theory may also be obtained by considering M-theory on $\CC \times \CC^2/\ZZ_2$. The $Y_N$ theories are obtained via orbifolds thereof.

\subsection{\texorpdfstring{$+_{N,M}$}{+-NM} solutions}

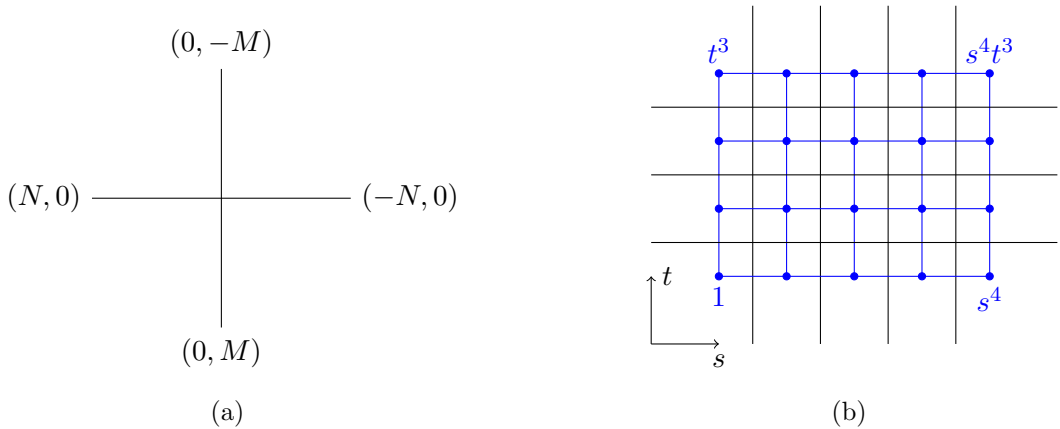
\begin{figure}
\centering
\subfigure[][]{\label{fig:plus-junction}
 \begin{tikzpicture}[scale=0.86]
   \draw (-2,0) node [anchor=east] {\small $(N,0)$} -- (2,0) node [anchor=west] {\small $(-N,0)$};
   \draw (0,-2) node [anchor=north] {\small $(0,M)$} -- (0,2) node [anchor=south] {\small $(0,-M)$};
 \end{tikzpicture}
}
\hskip 20mm
\subfigure[][]{\label{fig:Plus-grid}
 \begin{tikzpicture}[scale=0.9]
   \foreach \i in {-1,...,3}{
   \foreach \j in {0,...,3}{
     \draw[fill,blue] (3-\i,\j) circle (1.5pt);
   }
   \draw[blue] (3-\i,0) -- (3-\i,3);
   }
   \foreach \i in {-1,...,2}{
    \draw (3-\i-0.5,-1) -- (3-\i-0.5,4);
   }
   \foreach \i in {1,...,3}{
    \draw (-1,\i-0.5) -- (5,\i-0.5);
   }
   
   \foreach \j in {0,...,3}{
     \draw[blue] (0,\j) --(4,\j);
   }
   
   \draw[blue] (0,0)node [anchor=north] {\small $1$};
   \draw[blue] (0,3)node [anchor=south] {\small $t^3$};
   \draw[blue] (4,0)node [anchor=north] {\small $s^4$};
   \draw[blue] (4,3)node [anchor=south] {\small $s^4 t^3$};
   
    \draw[->] (-1,-1) -- (-1,0) node [anchor=west] {\small $t$};
    \draw[->] (-1,-1) -- (0,-1) node [anchor=north] {\small $s$};
 \end{tikzpicture}
}

 \caption{Left: the 5-brane junction describing the $+_{N,M}$ SCFT. Right: brane web and grid diagram for a mass deformation of the $+_{3,4}$ theory (a complete triangulation of the grid diagram can be obtained by resolving the remaining brane intersections).
 }
\end{figure}

The next example is a quartic junction of $N$ D5-branes and $M$ NS5-branes, as shown in fig.~\ref{fig:plus-junction}.
This configuration has been discussed already in \cite{Aharony:1997bh}.
An example for the associated grid diagram is shown in fig.~\ref{fig:Plus-grid}.
The polynomial $P(s,t)$ defining the M-theory curve is given by
\begin{align}\label{eq:Pst-plus-NM}
 P(s,t)&=\sum_{i=0}^M\sum_{j=0}^N c_{i,j} s^i t^j~.
\end{align}
The boundary conditions in the conformal limit are,
\begin{align}
 P(s,t)\big\vert_{s\rightarrow\infty, \text{$t$ finite}}&\stackrel{!}{\sim}  s^M(t-\alpha_1)^N~,
 &
 P(s,t)\big\vert_{ \text{$s$ finite}, t\rightarrow\infty}&\stackrel{!}{\sim} t^N(s-\alpha_2)^M~,
 \nonumber\\
 P(s,t)\big\vert_{\text{$s$ finite}, t\rightarrow 0}&\stackrel{!}{\sim}  (s-\alpha_3)^M~, 
 &
 P(s,t)\big\vert_{s\rightarrow 0, \text{$t$ finite}}&\stackrel{!}{\sim}(t-\alpha_4)^N~,
 \label{eq:plus-NM-bc}
\end{align}
with $|\alpha_i|=1$. Consistency of the boundary conditions requires $\alpha_1 \alpha_3 = \alpha_2 \alpha_4$.

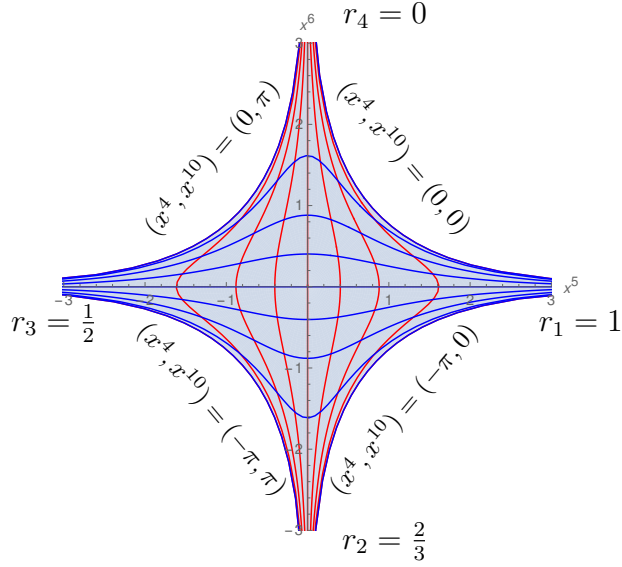
\begin{figure}
\centering
 \begin{tikzpicture}
  \node at (0,0) {\includegraphics[width=0.45\linewidth]{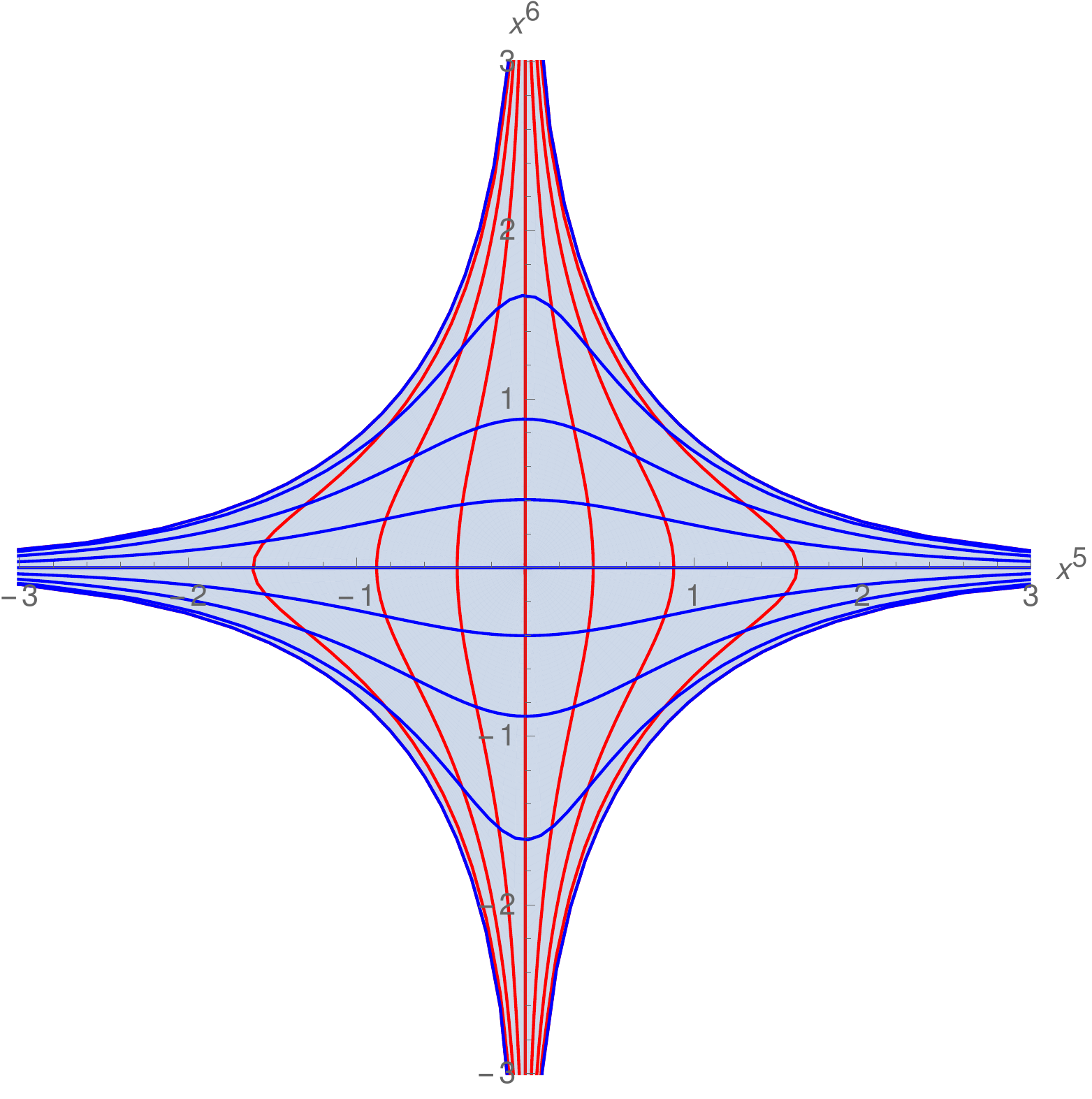}};
  \node at (0.9,-3.5) {$r_2=\frac{2}{3}$};
  \node at (0.9,3.5) {$r_4=0$};
  \node at (3.5,-0.6) {$r_1=1$};
  \node at (-3.5,-0.6) {$r_3=\frac{1}{2}$};
  
  \node[rotate=50] at (1.15,-1.75) {\footnotesize $(x^4,x^{10})=(-\pi,0)$};
  \node[rotate=50] at (-1.4,1.6) {\footnotesize $(x^4,x^{10})=(0,\pi)$};
  \node[rotate=-50] at (1.15,1.6) {\footnotesize $(x^4,x^{10})=(0,0)$};
  \node[rotate=-50] at (-1.4,-1.75) {\footnotesize $(x^4,x^{10})=(-\pi,\pi)$};
 \end{tikzpicture}
 \caption{
 $+_{1,1}$ curve with $\tilde R_4=R_{10}=1$ obtained by embedding $\Sigma_{\rm IIB}$ into $\cM_4$ via (\ref{eq:cApm-M4}). The poles $r_\ell$ on $\Sigma_{\rm IIB}$ correspond to the external 5-branes as indicated.  The segments of $\partial\Sigma_{\rm IIB}$ in between poles are mapped to the outer curves connecting the asymptotic regions, with $x^4$, $x^{10}$ as indicated. The blue and red curves correspond to constant $x^{4}$ and $x^{10}$, respectively. The embedding of the second half of $\hat\Sigma_{\rm IIB}$, with $w$ in the lower half-plane, is obtained via (\ref{eq:SigmaIIB-to-cc}).
\label{fig:Plus-1-curve}}
\end{figure}

We again show that the functions $\cA_\pm$ of the corresponding supergravity solution provide an explicit parametrization of the curve. They are given by (sec.~4.2 of \cite{Bergman:2018hin})
\begin{align}
 \cA_\pm&=\frac{3}{4}\alpha^\prime \left[\pm M(\ln(3w-2)-\ln w)+i N(\ln(2w-1)-\ln(w-1))\right]~.
\end{align}
From (\ref{eq:cA-sttilde}), $\s$ and $\t$ are obtained as
\begin{align}
 \s&=\left(\frac{2w-1}{w-1}\right)^N~, & \t&=\left(\frac{3w-2}{w}\right)^M~.
\end{align}
They satisfy
\begin{align}\label{+NMcurve}
 0&=\cP_{+_{N,M}}(\s,\t)~, & \cP_{+_{N,M}}(\s,\t)\equiv1+\s^{1/N}+\t^{1/M}-\s^{1/N}\t^{1/M}~.
\end{align}
To compare to the definition of the curve via (\ref{eq:Pst-plus-NM}), this equation can again be recast in terms of a polynomial $\tilde P_{+_{N,M}}(\s,\t)$. Namely,
\begin{align}
\label{eq:plus-polynomial}
 0&=\tilde P_{+_{N,M}}(\s,\t)~, &
 \tilde P_{+_{N,M}}(\s,\t)&\equiv\prod_{n=0}^{N-1}\prod_{m=0}^{M-1} \cP_{+_{1,1}}\left(e^{\frac{2 \pi i n}{N}} \s^{\frac{1}{N}}, e^{\frac{2 \pi i m}{M}} \t^{\frac{1}{M}} \right)~.
\end{align}
This indeed yields polynomials of the form (\ref{eq:Pst-plus-NM}) satisfying the boundary conditions in (\ref{eq:plus-NM-bc}). 
Some explicit examples for small $N$ are
\begin{align}
 \tilde c_{ij}^{+_{1,4}}&=\begin{pmatrix}
 1 & -1 \\
 4 & 4 \\
 6 & -6 \\
 4 & 4 \\
 1 & -1 \\
\end{pmatrix}~,
&
\tilde c_{ij}^{+_{5,3}}&=
\begin{pmatrix}
 1 & 5 & 10 & 10 & 5 & 1 \\
 3 & -495 & 3390 & -3390 & 495 & -3 \\
 3 & 495 & 3390 & 3390 & 495 & 3 \\
 1 & -5 & 10 & -10 & 5 & -1 \\
\end{pmatrix}~.
\end{align}
The binomial form of the edge coefficients again implies that the correct boundary conditions are satisfied.
The curve obtained from the supergravity solution is shown in fig.~\ref{fig:Plus-1-curve}.

The $+_{1,1}$ theory may also be obtained by considering M-theory on the conifold $\cC$. The $+_{N,M}$ theories are obtained by considering M-theory on $\cC / (\ZZ_N \times \ZZ_M)$, with the orbifold action given in (\ref{orbifoldaction}), but with $\nu^M=1$.

\subsection{\texorpdfstring{$X_{N,M}$}{X-N-M} solutions}

\begin{figure}
\centering
\subfigure[][]{\label{fig:X-NM-junction}
 \begin{tikzpicture}[scale=0.86]
   \draw (-2,2) node [anchor=south] {\small $(N,-N)$} -- (2,-2) node [anchor=north] {\small $(-N,N)$};
   \draw (-2,-2) node [anchor=north] {\small $(M,M)$} -- (2,2) node [anchor=south] {\small $(-M,-M)$};
 \end{tikzpicture}
}
\hskip 20mm
\subfigure[][]{\label{fig:X-grid}
 \begin{tikzpicture}[scale=0.9]
  \begin{scope}[rotate=45]
   \foreach \i in {-1,...,3}{
   \foreach \j in {0,...,3}{
     \draw[fill,blue] (3-\i,\j) circle (1.5pt);
   }
   \draw[blue] (3-\i,0) -- (3-\i,3);
   }
   \foreach \i in {-1,...,2}{
    \draw (3-\i-0.5,-1) -- (3-\i-0.5,4);
   }
   \foreach \i in {1,...,3}{
    \draw (-1,\i-0.5) -- (5,\i-0.5);
   }
   
   \foreach \j in {0,...,3}{
     \draw[blue] (0,\j) --(4,\j);
   }
   
   \draw[blue] (0,0)node [anchor=north] {\small $s^3$};
   \draw[blue] (0,3)node [anchor=east] {\small $t^3$};
   \draw[blue] (4,0)node [anchor=west] {\small $s^7 t^4$};
   \draw[blue] (4,3)node [anchor=south] {\small $s^4 t^7$};
 \end{scope}
 \begin{scope}[xshift=-20mm,yshift=5mm]
    \draw[->] (-1,-1) -- (-1,0) node [anchor=west] {\small $t$};
    \draw[->] (-1,-1) -- (0,-1) node [anchor=north] {\small $s$};
 \end{scope}
 \end{tikzpicture}
}

 \caption{Left: the 5-brane junction describing the $X_{N,M}$ SCFT. Right: brane web and grid diagram for a mass deformation of the $X_{4,3}$ theory (a complete triangulation of the grid diagram can be obtained by resolving the remaining brane intersections).
 }
\end{figure}
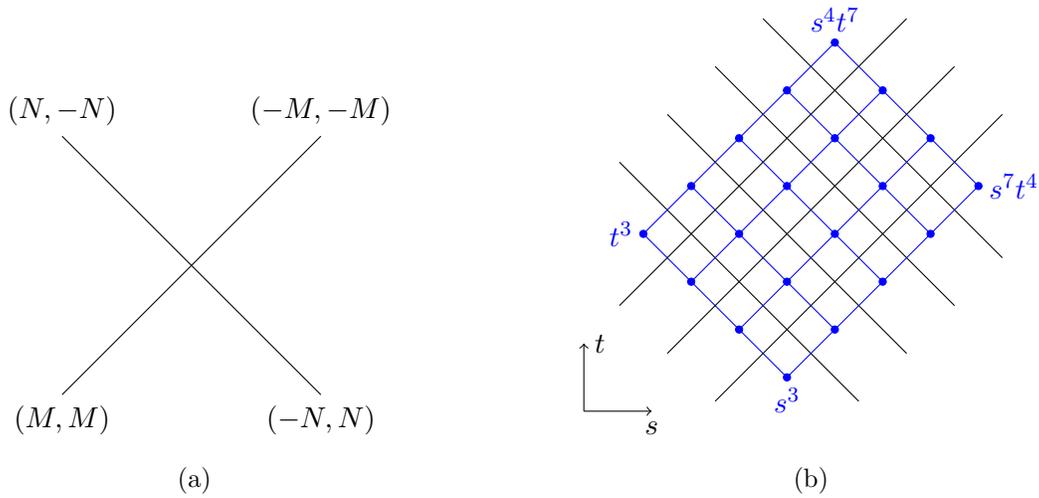

The $X_{N,M}$ theories are defined by quartic junctions of $N$ $(1,-1)$ 5-branes and $M$ (1,1) 5-branes, as in fig.~\ref{fig:X-NM-junction}. They are closely related to the $+_{N,M}$ theories, in a very similar way to how the $Y_N$ theories are related to the $T_N$ theories.

The quantities $\s$ and $\t$ extracted from the supergravity solution (as discussed in sec.~4.2.2 of \cite{Bergman:2018hin}) via (\ref{eq:cA-sttilde}) are
\begin{align}
\s&=\left(\frac{2w-1}{w-1}\right)^N\left(\frac{3w-2}{w}\right)^M~, & \t&=\left(\frac{3w-2}{w}\right)^M\left(\frac{w-1}{2w-1}\right)^N~.
\end{align}
These are related to the complex coordinates of the $+_{N,M}$ theory by
\begin{align}
\label{X+relation}
\s_{+_{N,M}} &= \sqrt{\frac{\s}{\t}}\Big|_{X_{N,M}} ~, & \t_{+_{N,M}} &=\sqrt{\s\, \t}~ \big|_{X_{N,M}} ~.
\end{align}
In Type IIB, the two configurations are related by an $SL(2,\RR)$ rotation, together with a rescaling of charges. However, the two configurations are not related by $SL(2,\ZZ)$ in the full string theory description, as can be seen by comparing (\ref{X+relation}) to (\ref{eq:st-SL2Z}). 
Using (\ref{X+relation}) and (\ref{+NMcurve}), the M-theory curve for the $X_{N,M}$ theory is 
\begin{align}
0&=\cP_{X_{N,M}}(\s,\t)~,
&
\cP_{X_{N,M}}(\s,\t)&\equiv\s^{\frac{1}{2 N}} + \t^{\frac{1}{2 N}} + (\s\t)^{\frac{1}{2M}}\left(\t^{\frac{1}{2 N}}- \s^{\frac{1}{2 N}}\right)~.\label{eq:P-XNM}
\end{align}
The factors of $1/2$ in the exponents imply that the $X_{N,M}$ M5-brane has twice the winding along the torus as the $+_{N,M}$ M5-brane. 

One can once again convert (\ref{eq:P-XNM}) to polynomial form, $\tilde P_{X_{N,M}}=0$. However, unlike for the $T_N$, $Y_N$, and $+_{N,M}$ curves, the grid diagram is not obtained by simply subdividing the lattice in the horizontal and vertical directions. Consequently, the polynomial for the general $X_{N,M}$ solutions does not follow the pattern in (\ref{eq:Ptilde-TN}), (\ref{eq:Y-polynomial}), (\ref{eq:plus-polynomial}).
Some examples for small $N$, $M$ are
\begin{align}
 \tilde c_{ij}^{X_{1,2}}&=
\begin{pmatrix}
  &  & 1 &  \\
  & -2 & 8 & -1 \\
 1 & 8 & 2 &  \\
  & -1 &  &  \\
\end{pmatrix}~,
&
\tilde c_{ij}^{X_{4,2}}&=
\begin{pmatrix}
  &  & 1 &  &  &  &  \\
  & -2 & -128 & -4 &  &  &  \\
 1 & -128 & 2568 & -1920 & 6 &  &  \\
  & -4 & -1920 & -13324 & -1920 & -4 &  \\
  &  & 6 & -1920 & 2568 & -128 & 1 \\
  &  &  & -4 & -128 & -2 &  \\
  &  &  &  & 1 &  &  \\
\end{pmatrix}~.
\end{align}
These are generally polynomials of precisely the form implied by the grid diagram (fig.~\ref{fig:X-grid}), with binomial edge coefficients implementing the boundary conditions.

The $X_{1,1}$ theory may be described as M-theory on the cone over $\mathds{F}^0 = {\mathds{P}}^1 \times {\mathds{P}}^1$.

\subsection{\texorpdfstring{$\pslash_{N}$}{pslash-N} solutions}

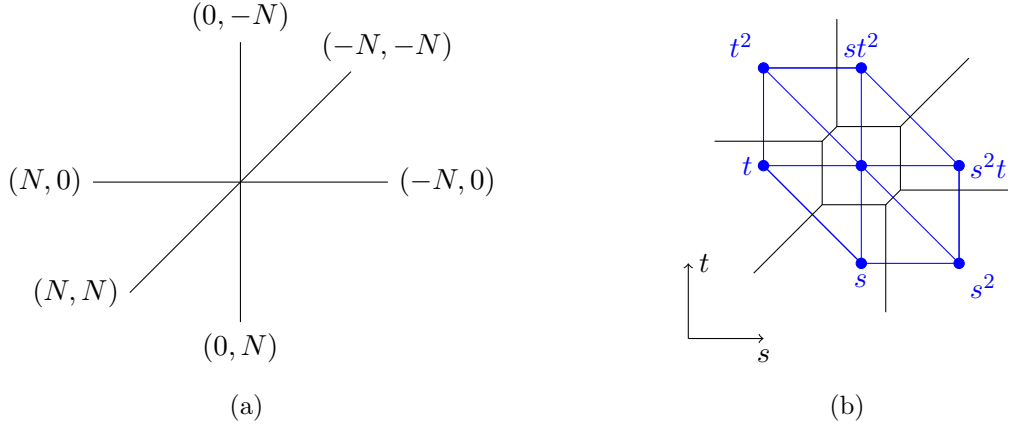
\begin{figure}
\centering
\subfigure[][]{
 \begin{tikzpicture}[scale=0.98]\label{eq:pslash-junction}
   \draw (-1,0) node [anchor=east] {\small $(N,0)$} -- (3,0) node [anchor=west] {\small $(-N,0)$};
   
  \draw (1,-1.9)node [anchor=north] {\small $(0,N)$} -- (1,1.9) node [anchor=south] {\small $(0,-N)$};
   \draw (-0.5,-1.5) node [anchor=east] {\small$(N,N)$ }-- (1,0)-- (2.5,1.5) node [anchor=south] {\small \,\,\,\,\,\,\,\,\,\,\,\,\,\,\,$(-N,-N)$};
 \end{tikzpicture}
}
\hskip 20mm
\subfigure[][]{
 \begin{tikzpicture}
 \begin{scope}[scale=1.3]

    \draw[fill,blue] (1,0) circle (1.5pt) node[anchor=north] {\small $s^{}$};
    \draw[fill,blue] (2,0) circle (1.5pt)node[anchor=north west] {\small $s^2$};
    \draw[fill,blue] (2,1) circle (1.5pt)node[anchor=west] {\small $s^2 t$};
    \draw[fill,blue] (1,2) circle (1.5pt) node[anchor=south] {\small $s t^2$};
    \draw[fill,blue] (0,2) circle (1.5pt) node[anchor = south east] {\small $t^2$};
    \draw[fill,blue] (0,1) circle (1.5pt) node[anchor =  east] {\small $t$};
    \draw[fill,blue] (1,1) circle (1.5pt);

    \draw[blue] (1,0) -- (2,0) -- (2,1) -- (1,2) -- (0,2)--(0,1) -- (1,0);
    \draw[blue] (0,2) -- (2,0);
    \draw[blue] (0,1) -- (1,0);
    \draw[blue] (0,2) -- (1,2);
    \draw[blue] (0,1) -- (2,1);
    \draw[blue] (1,0) -- (1,2);
    \draw[blue] (2,0) -- (2,1);

 %  \draw (-0.5,0.25) -- (0.25,0.25) -- (0.25,-0.5);
   \draw (-0.1,-0.1) -- (0.6,0.6) -- (1.25,0.6) -- (1.25,-0.5);
   \draw (0.6,0.6) -- (0.6,1.25) -- (-0.5,1.25);
   
   \draw (0.6,1.25) -- +(0.15,0.15) -- (0.75,2.5);
   %\draw (0.75,2.15) -- +(0.5,0.5);
   
   \draw (1.25,0.6) -- +(0.15,0.15)--(2.5,0.75) ;
   %\draw (2.15,0.75) -- +(0.5,0.5);
   
   \draw (0.75,1.4) -- (1.4,1.4) -- (1.4,0.75);
   \draw (1.4,1.4) -- +(0.7,0.7);
   \end{scope}
    \draw[->] (-1,-1) -- (-1,0) node [anchor=west] {\small $t$};
    \draw[->] (-1,-1) -- (0,-1) node [anchor=north] {\small $s$};
 \end{tikzpicture}
}
 \caption{Left: the sextic junction describing the $\pslash_N$ theory. Right: 
 brane web and grid diagram for a deformation of the $\pslash_1$ theory.\label{fig:PlusSlash-grid}}
\end{figure}

As a final example we consider the $\pslash_{N}$ theories, which are realized by sextic junctions of NS5, D5, and (1,1) 5-branes as shown in fig.~\ref{eq:pslash-junction}. The polynomial $P(s,t)$ obtained from the grid diagram takes the form
\begin{align}\label{eq:pslash-poly}
P(s,t) = \sum_{\substack{0 \leq i\,,\,j \leq 2 N \\ N \leq i + j \leq 3 N}} c_{i,j}s^i t^j~.
\end{align}
The boundary conditions are
\begin{align}
 P(s,t)\big\vert_{s,t\rightarrow\infty}&\stackrel{!}{\sim}   s^N t^N (s-\alpha_1 t)^N~,
 &
 P(s,t)\big\vert_\text{$s,t\rightarrow 0$}&\stackrel{!}{\sim} (s-\alpha_4 t)^N~,
 \nonumber\\
 P(s,t)\big\vert_\text{$t$ finite, $s\rightarrow 0$}&\stackrel{!}{\sim} t^N (t-\alpha_3)^N~, 
 &
 P(s,t)\big\vert_\text{$s$ finite, $t\rightarrow \infty$}&\stackrel{!}{\sim}   t^{2N} (s-\alpha_2)^N~,
 \nonumber\\
 P(s,t)\big\vert_\text{$s$ finite, $t\rightarrow 0$}&\stackrel{!}{\sim}s^{N} (s-\alpha_5)^N~,
 &
 P(s,t)\big\vert_\text{$t$ finite, $s\rightarrow \infty$}&\stackrel{!}{\sim}s^{2N} (t-\alpha_6)^N~,
 \label{eq:pslash-bc}
\end{align}
with $|\alpha_i|=1$. For consistency, we require that $\alpha_1\alpha_2 \alpha_3 = \alpha_4 \alpha_5 \alpha_6$.

The supergravity solution has been discussed in sec.~4.5 of  \cite{Bergman:2018hin}. Via (\ref{eq:cA-sttilde}), $\s$ and $\t$ are found to be
\begin{align}
\s&=\left(\frac{1}{\sqrt{7 + 4 \sqrt{3}} }\frac{(w-r_5)(w-r_6)}{(w-r_2)(w-r_3)} \right)^N, & 
\t&=\left(\sqrt{7 + 4 \sqrt{3}}\,\frac{(w-r_1)(w-r_6)}{(w-r_3)(w-r_4)} \right)^N,
\end{align}
where 
\begin{align}
r_1 &=-r_2= - 2 + \sqrt{3}~, &r_4 &=-r_5= 2 + \sqrt{3}~, & r_3 &=-r_6=1~.
\end{align}
They satisfy $\cP_{\pslash_N}(\s, \t)=0$ with
\begin{align}\label{eq:cP-pslash}
\cP_{\pslash_N}(\s, \t) &= \left(\s^{1/N} + \t^{1/N}\right)\left(1+(\s\t)^{1/N}\right) - \s^{2/N} - \t^{2/N}+6 (\s\t)^{1/N}~.
\end{align}
For $N=1$ this is a polynomial. Converting the equation for generic $N$ to polynomial form yields
\begin{align}
\tilde P_{\pslash_N}(\s,\t)&\equiv\prod_{n=0}^{N-1}\prod_{m=0}^{N-1} \cP_{\pslash_1}\left(e^{\frac{2 \pi i n}{N}} \s^{\frac{1}{N}}, e^{\frac{2 \pi i m}{N}} \t^{\frac{1}{N}} \right)~.
\end{align}
These are polynomials of the form (\ref{eq:pslash-poly}), satisfying the constraints spelled out in (\ref{eq:pslash-bc}).
This establishes the identification of Type IIB supergravity solutions with M-theory curves, (\ref{eq:st-sttilde}), also for this class of solutions.
An example polynomial is
\begin{align}
 \tilde c_{ij}^{\pslash_3}&=\begin{pmatrix}
  &  &  & 1 & -3 & 3 & -1 \\
  &  & 3 & 2172 & 9474 & 2172 & 3 \\
  & 3 & -9474 & 400119 & -400119 & 9474 & -3 \\
 1 & 2172 & 400119 & 2444568 & 400119 & 2172 & 1 \\
 -3 & 9474 & -400119 & 400119 & -9474 & 3 &  \\
 3 & 2172 & 9474 & 2172 & 3 &  &  \\
 -1 & 3 & -3 & 1 &  &  &  \\
\end{pmatrix}
~.
\end{align}

The $\pslash_{1}$ theory may be obtained from M-theory on the cone over dP$_3$. The $\pslash_{N}$ theory is obtained by a $\ZZ_N \times \ZZ_N$ orbifold of this geometry.

\begin{acknowledgments}
We thank Barak Kol, Philip Argyres and Fengjun Xu for helpful discussions.
Part of this work was completed during the workshop ``Strings, Branes and Gauge Theories'' at APCTP;
CFU thanks the organizers for the interesting workshop and APCTP for hospitality and support.
Final stages of this work were performed at the Aspen Center for Physics, which is supported by National Science Foundation grant PHY-1607611. CFU thanks the organizers of the program ``Superconformal Field Theories and Geometry'' for the interesting workshop.
JK would like to thank the Yukawa Institute for Theoretical Physics and the Simons Center for Geometry and Physics for their hospitality during the completion of this work.
This work is supported in part by the National Science Foundation under grant PHY-16-19926.  
\end{acknowledgments}

\bibliographystyle{JHEP}
\bibliography{ads6}
\end{document}